\documentstyle[12pt,epsf,aasms4]{article}

\def\part#1#2{\frac{\partial #1}{\partial #2}}
\def\xb{\bar\xi(a,x)}
\def\gaprox{\mbox{$\,$ 
\raisebox{0.5ex}{$<$}\hspace{-1.7ex}{\raisebox{-0.5ex}{$\sim$ }}$\,$} }

\begin{document}

\title{Nonlinear gravitational clustering : dreams of a paradigm}
\author{T.Padmanabhan \& Sunu Engineer}
\affil{Inter University Centre for Astronomy and Astrophysics \protect\\
         Postbag 4, Ganeshkhind, Pune 411007 (India).\\
	 E-mail: paddy@iucaa.ernet.in : sunu@iucaa.ernet.in}

\abstract
We discuss the {\it late time} evolution of the gravitational clustering in
an expanding universe, based on the nonlinear scaling relations (NSR) which
connect the nonlinear and linear two point correlation functions. The existence
of critical indices for the NSR suggests that the evolution may proceed towards
a universal profile which does not change its shape at late times. We begin
 by clarifying the relation between the density profiles of the individual halos and
the slope of the correlation function and discuss
the conditions under which the slopes of the correlation function at the
extreme nonlinear end can be independent of the initial power spectrum.
 If the
evolution should lead to a profile which preserves the shape at late times,
then the correlation function should grow as $a^2$ [in a $\Omega=1$ universe] even at nonlinear scales.
We prove that such {\it exact} solutions do not exist; however, there exists a 
class of solutions (``psuedo-linear profiles'', PLP's for short) which evolve as $a^2$ to a good approximation. It turns out that the PLP's are the correlation
functions which arise if the individual halos are
assumed to be isothermal spheres. They are also configurations of mass in
which the nonlinear effects of gravitational clustering is a minimum and hence
can act as building blocks of the nonlinear universe. We discuss the implications of this result. 

\section{Introduction}

The evolution of  large number of particles under their mutual
gravitational influence is a well-defined mathematical problem. If such a
system occupies a finite region of phase space at an initial instant, and evolves
via newtonian gravity, then it does not reach any sensible `equilibrium' 
state. The core region of the system will keep on shrinking and will be
eventually be dominated by a few hard binaries. Rest of the particles will
evaporate away to large distances, gaining kinetic energy from the shrinking
core [for a discussion of  such systems, see \cite{PadPhyRep}].

The situation is drastically different in the presence of an expanding 
background universe characterised by an expansion factor $a(t)$. Firstly, the 
expansion tends to keep particles apart 
thereby exerting a civilising influence against newtonian attraction. Secondly,
it is now possible to consider an infinite region of space filled with 
particles.
 The average density of particles will contribute to the expansion of the 
background universe and the deviations from the uniformity will lead to
clustering. Particles evaporating from a local overdense cluster cannot
escape to ``large distances'' but necessarily will encounter other deep
potential wells. Naively, one would expect the local overdense regions to 
eventually 
form gravitationally bound objects, with a hotter distribution of particles
hovering uniformly all over. As the background expands, the 
velocity dispersion of the second component will keep decreasing and they will
be captured  by the deeper potential wells. Meanwhile, the clustered component 
will also evolve dynamically and participate in, e.g mergers. If the background 
expansion and  the initial
conditions have no length scale, then it is likely that the clustering  will 
continue
in a hierarchical manner {\it ad infinitum}.
\par

Most of the practising cosmologists will broadly agree with
the above  picture of gravitational clustering in an expanding universe. It is, 
however, not
easy to translate these concepts into a well-defined mathematical formalism
and provide a more quantitative description of the gravitational clustering.
One of the key questions regarding this system which needs to be addressed is 
the following: Can one make any general statements about the very late stage
evolution of the clustering ? For example, does the power spectrum at late times
`remember' the initial power spectrum or does it possess some universal
characteristics which are reasonably independent of initial conditions ?
[This question is closely related to the issue of whether gravitational 
clustering leads to density profiles which are universal.   
\cite{Navarro}].

\par
We address some aspects of this issue in this paper and show that it is possible to 
provide (at least partial) answers to  these questions based on a simple 
paradigm.
The key assumption we shall make is the following: Let ratio 
 between
mean relative pair velocity $v(a,x)$ and the negative hubble velocity ($-\dot a 
x$) 
be denoted by $h(a,x)$ and let $\bar\xi(a,x)$ be the mean correlation
function averaged over a sphere of radius $x$. {\it We shall assume that
$h(a,x)$ depends on $a$ and $x$ only through   
$\bar\xi(a,x)$; that is, $h(a,x)=h[\bar\xi(a,x)]$.} With such  a minimal 
assumption, we will be able to obtain several conclusions regarding the
evolution of power spectrum in the  universe. Such an assumption was originally
introduced --- in a different form --- by Hamilton (\cite{Ham}).
The present form, as well as its theoretical implications were discussed
in  \cite{RajPad94}, and a theoretical model for the
scaling was attempted by Padmanabhan (\cite{TPMNRAS}). 
It must be noted that 
simulations indicate a dependence of the relation $h(a,x)=h[\bar\xi(a,x)]$
on the intial spectrum and also on cosmological parameters [\cite{PeaDodd};
 \cite{PeaDodd96}; \cite{PadOst}; \cite{MoJain}]. Most of our discussion in this
paper is independent of this fact or can be easily generalised to such cases.
When we need to use an explicit form for $h$ we shall use the original
ones suggested by Hamilton (\cite{Ham}) because of its simplicity. 

Since this paper addresses several independent but related  questions,
we provide here a brief summary of how it is organised. Section 2 studies
some aspects of nonlinear evolution based on the assumption mentioned above.
We begin by summarising some previously known results in subsection 2.1 to
set up the notation and collect together in one place the formulas we need later.
Subsection 2.2 makes a brief comment about the critical indices in gravitational
dynamics
so as to motivate later discussion. In section 3, we discuss the relation between density profiles
of halos and correlation functions and derive the conditions under which one
may expect universal density profiles in gravitational clustering. In section
4 we show that gravitational clustering does {\it not} admit self similar
evolution except in a very special case. We also discuss the conditions for
approximate self-similarity to hold. Section 5 discusses the question as whether
one can expect to find power spectra which evolve preserving their shape, even in the nonlinear
regime. We first show, based on the results of section 4, that such {\it exact}
solutions cannot exist. We then discuss the conditions for the existence of
some {\it approximate} solutions. We obtain one prototype approximate solution
and discuss its properties. The solution also allows us to understand the
connection between statistical mechanics of gravitating systems in the small
scale and evolution of correlation functions on the large scale. Finally,
section 6 discusses the results. 
\vskip 0 pt plus -20pt
\section{General features of nonlinear evolution}
Consider the evolution of the system 
starting from a gaussian initial fluctuations with an initial power
spectrum, $P_{in}(k)$. The fourier transform of the power spectrum 
defines the correlation function $\xi(a,x)$ where $a\propto t^{2/3}$
is the expansion 
factor in a universe with $\Omega=1$. It is more convenient to work with the
average correlation function inside a sphere of radius $x$, 
defined by

\begin{equation}
\label{avxi}
\bar{\xi}(a,x)\equiv {3\over x^3}\int^{x}_{0}\xi(a,y)\: y^2 dy  
\end{equation}
This quantity is related to the power spectrum $P(a,k)$ by
\begin{equation}
\bar\xi(x,a)=\frac{3}{2 \pi^2 x^3} \int_{0}^{\infty} \frac{dk}{k} P(a,k)\left[ 
\sin(k\:x)-k\:x\:\cos(k\:x)\right]
\end{equation}
with the inverse relation
\begin{equation}
P(a,k)=\frac{4 \pi}{3 k}\int_{0}^{\infty} dx\: x\: \bar\xi(a,x)\left[ 
\sin(k\:x)-k\:x\:\cos(k\:x)\right]
\end{equation}
In the linear regime we have $\bar{\xi}_L(a,x)\propto
a^2\bar\xi_{in}(a_i,x)$.
\par
We now recall that the conservation of pairs of particles gives an exact
equation satisfied by the correlation function (\cite{Peeb80}):

\begin{equation}
\label{pairconserv}
{\partial\xi \over\partial t}+{1\over ax^2}{\partial\over\partial
x}[x^2(1+\xi)v]=0
\end{equation}
where $v(a,x)$ denotes the mean relative velocity of pairs at
separation $x$ and 
epoch $a$. Using the mean correlation function $\bar\xi$ and a
dimensionless pair velocity $h(a,x) \equiv - (v/\dot{a}x)$, equation
(\ref{pairconserv}) can be written as
\begin{equation}
\label{redefpair}
({\partial\over\partial \ln a}-h{\partial\over\partial \ln x})\,\,\,
(1+\bar{\xi})=3h(1+\bar{\xi})
\end{equation}
This equation can be simplified by introducing the variables
\begin{equation}
A=\ln a,\qquad X=\ln x ,\qquad D(X,A) = \ln (1+\bar{\xi})
\end{equation}
in terms of which we have (\cite{RajPad94})

\begin{equation}
\label{qkey}
\frac{\partial D}{\partial A}-h(A,X)\frac{\partial D}{\partial
X}= 3h(A,X)
\end{equation}
At this stage we shall introduce our key assumption, viz. that
$h$ depends on $(A,X)$ only through $\bar\xi$ (or, equivalently, $D$).
Given this single assumption, several results follow which we shall
now summarise.

\subsection{{\it Formal solution}}

Given that $h=h[\bar\xi(a,x)]$, one can easily integrate the equation
 (\ref{redefpair})
to find the general solution [see \cite{RajPad94} ]. The characteristics 
of this equation (\ref{redefpair}) satisfy the condition 
\begin{equation}
\label{qxandl}
x^3(1+\bar{\xi})=l^3
\end{equation}
where $l$ is another length scale. When the evolution is linear at all
the relevant scales, $\bar{\xi}\ll 1$ and $l\approx x$. As clustering
develops,
$\bar{\xi}$ increases and $x$ becomes considerably smaller than $l$.
The behaviour of clustering at some scale $x$ is then
determined by
the original {\it linear} power spectrum at the scale $l$ through the
``flow of information'' along the characteristics.
This suggests that {\it we can  express the true
correlation function $\bar\xi(a,x)$ in terms of the linear correlation
function $\bar\xi_L(a,l)$ evaluated at a different point}.
This is indeed true and the general solution can be expressed as
a nonlinear scaling relation (NSR, for short) between $\bar\xi_L(a,l)$ and $\bar\xi(a,x)$ with $l$ and $x$ related 
by equation(\ref{qxandl}). To express this solution we define
two functions ${\cal V}(z)$  and ${\cal U}(z)$ where ${\cal V}(z)$  is related 
to the function $h(z)$ by

\begin{equation}
{\cal V}(z)=\exp\left(\:\frac{2}{3} \int^{z} \frac{d z}{h(z)\,(1+z)} \right)
\end{equation} 
and
${\cal U}(z)$
is the inverse function of ${\cal V}(z)$. Then the solution to the
equation (\ref{redefpair}) can be written in either of two equivalent forms as:
\begin{equation}
\label{mapfun}
\bar\xi(a,x)={\cal U}\left[\bar\xi_L(a,l)\right];\qquad \bar\xi_L(a,l)={\cal 
V}\left[\bar \xi(a,x)\right]
\end{equation}
where $l^3=x^3 (1+\bar\xi)$
(\cite{RajPad94}). Given the form of $h(\bar\xi)$
this allows one to relate the nonlinear correlation function to the linear
one.

From general theoretical considerations [see \cite{TPMNRAS}] it can be shown 
that ${\cal V}(z)$ has the form:

\begin{equation}
\label{Vapprox}
{\cal V}(z)=\left\{
\begin{array}{ll}
1 & (z \ll 1) \\
z^{1/3} & (1 \gaprox z \gaprox 200) \\
z^{2/3} & (200 \ll z ) \\
\end{array}
\right. 
\end{equation}
In these three regions $h(z)\approx[(2z/3),2,1]$ respectively. We shall
call these regimes, linear, intermediate and nonlinear respectively. More exact fitting 
functions to ${\cal V}(z)$ and ${\cal U}(z)$ have been suggested in literature. 
[see \cite{Ham}; \cite{MoJain}; \cite{PeaDodd}]. When needed in this paper, we 
shall use the one  given in Hamilton et al.,1991:
\begin{equation}
\label{ham1}
{\cal V}(z)=z 
\left(\frac{1+0.0158\:z^2+0.000115\:z^3}{1+0.926\:z^2-0.0743\:z^3+0.0156\:z^4}
\right)^{1/3}
\end{equation}
\begin{equation}
\label{ham2}
{\cal U}(z)=\frac{z+0.358\:z^3+0.0236\:z^6}{1+0.0134\:z^3+0.0020\:z^{9/2}}
\end{equation}
Equations (\ref{mapfun}) and (\ref{ham1},\ref{ham2}) implicitly determine 
$\bar\xi(a,x)$ in terms of $\bar\xi_L(a,x)$.

\subsection{{\it Critical indices}}

These NSR already allow one to obtain some general conclusions regarding
the evolution. To do this most effectively, let us define a local 
index for rate of clustering by
\begin{equation}
n_a(a,x)\equiv \part{\ln \xb}{\ln a}
\end{equation}
which measures how fast $\xb$ is growing. When $\xb\ll 1$, then $n_a=2$
irrespective of the spatial variation of $\xb$ and the evolution preserves the shape of $\xb$. However, as clustering develops, the growth rate will
depend on the spatial variation of $\xb$. Defining the effective spatial
slope by
\begin{equation}
-[n_{eff}(a,x)+3]\equiv \part{\ln \xb}{\ln x}
\end{equation}
one can rewrite the equation (\ref{redefpair}) as
\begin{equation}
\label{naeqn}
n_a=h(\frac{3}{\xb} -n_{eff})
\end{equation}
At any given scale of nonlinearity, decided by $\xb$, there exists a critical
spatial slope $n_c$ such that $n_a>2 $ for $n_{eff}<n_c$ [implying rate of growth is faster
than predicted by linear theory] and 
$n_a<2 $ for $n_{eff}>n_c$ [with the rate of growth being slower
than predicted by linear theory]. The critical index is fixed by setting $n_a=2$ in  equation (\ref{naeqn}) at any instant. This feature will tend  to ``straighten out" correlation functions  towards the critical slope.
[We are assuming that $\xb$ has a slope that is decreasing with
scale, which is true for any physically interesting case]. From the fitting function it is easy to see that in the range $1 {\mbox{\gaprox}} 
\bar\xi {\mbox{\gaprox}} 200$, the critical index is $n_c\approx -1$
and for $200 \gaprox \bar\xi$, the critical index is $n_c\approx -2$ (\cite{JsbTp}).
This clearly suggests that the local effect of evolution is to
drive the correlation function to have a shape with $(1/x)$ behaviour
at nonlinear regime and $(1/x^2)$ in the intermediate regime. Such a 
correlation function will have $n_a\approx 2$ and hence will grow at
a rate close to $a^2$. We shall say more about
this in section 3 below.
 
\section{ Correlation functions, density profiles and stable clustering}

Now that we have a NSR giving  $\xb$ in terms of $\bar\xi_L(a,l)$  
we can ask the question:
How does $\xb$ behave at highly nonlinear scales or, equivalently, at any
given scale at large $a$ ? 
\par
To begin with, it is easy to see that we must have $v=-\dot a x$ or  $h=1$ for 
sufficiently large $\bar\xi(a,x)$ {\it if we assume} that the
evolution gets frozen in proper coordinates at highly nonlinear scales. 
Integrating equation (\ref{redefpair}) with $h=1$, we get $\bar\xi(a,x)=a^3 F(ax)$;
we shall call this phenomenon ``stable clustering''. There are two points
which need to be emphasised about stable clustering:
\par
(1) At present, there exists some evidence 
from simulations (\cite{PadOst}) that 
stable clustering does {\it not} occur in a $\Omega=1$ model. In a {\it formal} sense, numerical simulations cannot disprove [or
even prove, strictly speaking] the occurrence of stable clustering, because of the finite dynamic
range of any simulation. 

(2). Theoretically speaking, the ``naturalness'' of stable clustering is
often overstated. The usual argument is based on the assumption that
at very small scales --- corresponding to high nonlinearities --- the structures
are ``expected to be" frozen at the proper coordinates. However, this argument does not
take into account the fact that mergers are not negligible at {\it any scale} in
an $\Omega=1$ universe. In fact,  stable clustering
is more likely to be valid in models with $\Omega<1$ --- a claim which seems to 
be again supported by simulations (\cite{PadOst}).

{\it If} stable clustering {\it is} valid, then the late time  behaviour of $\xb$ 
{\it cannot}
be independent of initial conditions. In other words the two requirements:
(i) validity of stable clustering at highly nonlinear scales and
(ii) the independence of late time behaviour from initial conditions, 
are mutually
exclusive. This is most easily seen for initial power spectra which
are scale-free. If $P_{in}(k)\propto k^n$ so that $\bar\xi_L(a,x)\propto a^2 
x^{-(n+3)}$, then it is
easy to show that $\xb$ at small scales will vary as
\begin{equation}
\bar\xi(a,x) \propto a^{\frac{6}{n+5}} x^{-\frac{3(n+3)}{n+5}};\qquad (\bar\xi 
\gg 200)
\end{equation}
if stable clustering is true. Clearly, the power law index in the nonlinear 
regime ``remembers''
the initial index. The same result holds for more general initial conditions.

What does this result imply for the profiles of individual halos?
To answer this question, let us start with the simple assumption that the density field $\rho(a,{\bf x})$ at late stages  can 
be expressed as a superposition
of several halos, each with some density profile; that is, we take
\begin{equation}
\label{haloes}
\rho(a,{\bf x})=\sum_{i} f({\bf x}-{\bf  x}_i,a)
\end{equation}
where the $i$-th halo is centered at ${\bf x}_i$ and contributes
an amount $f({\bf x}-{\bf  x}_i,a)$  at the location ${\bf x}_i$  [We can easily generalise this equation to the situation in which there are halos with
different properties, like core radius, mass etc by summing over the number
density of objects with particular properties; we shall not bother to
do this. At the other extreme, the exact description merely corresponds to taking
the $f$'s to be Dirac delta functions]. The power spectrum for the 
density contrast, $\delta(a,{\bf x})=(\rho/\rho_b-1)$, corresponding to the 
$\rho(a,{\bf x})$ in (\ref{haloes})  can be expressed as
\begin{eqnarray}
\label{powcen}
P({\bf k},a) &\propto& \left( a^3 \left| f({\bf k},a)\right| \right)^2 \left| 
\sum_i \exp -i {\bf k}\cdot{\bf x}_i(a) \right|^2   \\
\label{powcen1}
& \propto & \left( a^3 \left| f({\bf k},a)\right| \right)^2 P_{\rm cent}({\bf 
k},a)
\end{eqnarray}
 where $P_{\rm cent}({\bf k},a)$
denotes the power spectrum of the distribution of centers of the halos.
\par

If  stable clustering is valid, then the density profiles of halos are
frozen in proper coordinates and we will have $f({\bf x} -{\bf x}_i,a)=
f(a\:({\bf x}-{\bf x}_i))$;
hence the fourier transform will have the form $f({\bf k},a)=f({\bf k}/a)$. On 
the other
hand, the power spectrum at scales which participate in stable clustering
must satisfy $P({\bf k},a)=P({\bf k}/a)$ [This is merely the requirement 
$\xb=a^3F(ax)$
re-expressed in fourier space]. From equation (\ref{powcen1}) it follows that we 
must have
$P_{\rm cent}({\bf k},a)={\rm constant} $ independent of ${\bf k}$ and $a$ at 
small length scales. This can arise in the special case of 
random distribution of centers or --- more importantly --- because  we are 
essentially probing the interior of a single halo at sufficiently small scales. 
[Note that we must {\it necessarily} have $P_{\rm cent} \approx {\rm constant}$, for length scales smaller than typical halo size, by definition]. 
We can relate the halo profile to the correlation function
using  (\ref{powcen1}).  In particular, if the halo profile is a power law with 
$f\propto r^{-\epsilon}$,  it
follows that the $\xb$ scales as $x^{-\gamma}$ [ see also  \cite{silkmac}; \cite{SethJain}] where
\begin{equation}
\label{gammep}
\gamma=2\epsilon-3
\end{equation}
\par
Now if the {\it correlation function} scales as $[-3(n+3)/(n+5)]$, then
 we see that
the halo density profiles should be related to the initial power law
index through the relation 
\begin{equation}
\epsilon=\frac{3(n+4)}{n+5}
\end{equation} 
So clearly,  
the halos of
highly virialised systems still ``remember'' the initial power 
spectrum.
\par

Alternatively, one can try to ``reason out'' the profiles of the individual
halos and use it to obtain the scaling relation for correlation functions.
One of the favourite arguments used by cosmologists to obtain such a ``reasonable'' halo profile is based on spherical, scale invariant,
collapse.  It turns out
that one can provide a series of arguments, based on spherical collapse, to
show that --- under certain circumstances --- the {\it density profiles} at the
nonlinear end scale as $[-3(n+3)/(n+5)]$. The simplest variant of this argument
runs as follows: If we start with an initial density
profile which is $r^{-\alpha}$, then scale invariant spherical collapse
will lead to a profile which goes as $r^{-\beta}$ with $\beta=3\alpha/
(1+\alpha)$ [see eg., Padmanabhan, 1996, 1996a and references cited
therein]. Taking the intial slope
as $\alpha=(n+3)/2$ will immediately give $\beta=3(n+3)/(n+5)$. [Our definition of the stable clustering in the last section 
is based on the scaling of
the correlation function and gave the
slope of $[-3(n+3)/(n+5)]$ for the {\it correlation} function. The spherical
collapse gives the same slope for {\it halo profiles}.] In this case, when the halos have the slope of $\epsilon=3(n+3)/(n+5)$,
then the correlation function should have slope
\begin{equation}
\gamma=\frac{3(n+1)}{n+5}
\end{equation}
Once again, the final state ``remembers'' the initial index $n$.

Is this conclusion true ? Unfortunately, simulations do not have sufficient
dynamic range to provide a clear answer but there are some claims [see 
\cite{Navarro} ]  that
the halo profiles are ``universal'' and independent of initial conditions.
The theoretical arguments given above are also far from rigourous (in spite
of the popularity they seem to enjoy!). The argument for correlation function to scale as
$[-3(n+3)/(n+5)]$ is based on the assumption of $h=1$ asymptotically, which
may not be true. The argument, leading to density profiles scaling as
$[-3(n+3)/(n+5)]$, is based on scale invariant spherical collapse which
does not do justice to nonradial motions. Just to illustrate the situations
in which one may obtain final configurations which are independent of
initial index $n$, we shall discuss two possibilities:

(i) As a first example we will try to see when the slope of the correlation
function is universal and obtain the slope of halos in the nonlinear limit
using our relation (\ref{gammep}). Such an interesting situation can develop {\it if we assume that $h$ reaches a 
constant
value asymptotically which is not necessarily unity}. In that case, we can
integrate our equation (\ref{redefpair}) to get   $\xb=a^{3h}F[a^h x]$ where $h$ now
denotes the constant asymptotic value of of the function. For an initial
spectrum which is scale-free power law with index $n$, this result translates
to 
\begin{equation}
\bar\xi(a,x)\propto a^{\frac{2 \gamma}{n+3}} x^{-\gamma}
\end{equation} where $\gamma$ is given by 
\begin{equation}
\gamma=\frac{3 h (n+3)}{2+h(n+3)}
\end{equation}
We now notice that one can obtain
a $\gamma$  which is independent of initial power law index provided
$h$ satisfies the condition $h(n+3)=c$, a constant.  In this case, the nonlinear 
correlation
function will be given by
\begin{equation}
\bar\xi(a,x)\propto a^{\frac{6c}{(2+c)(n+3)}} x^{-\frac{3c}{2+c}}
\end{equation} 
The halo index will be independent of $n$
and will be given by
\begin{equation}
\epsilon=3\left( \frac{c+1}{c+2} \right) 
\end{equation}
Note that we are now demanding the asymptotic value of $h$ to {\it explicitly 
depend} on the initial conditions though the {\it spatial} dependence of $\xb$ 
does not.
In other words, the velocity distribution --- which is related to $h$ --- still 
``remembers'' the initial
conditions. This is indirectly reflected in the fact that the growth
of $\xb$ --- represented by $a^{6c/((2+c)(n+3))}$ --- does depend on the
index $n$.

\par
As an example of the power of such a --- seemingly simple --- analysis note the 
following: Since $c \geq 0 $, it follows that $\epsilon > (3/2)$; invariant 
profiles
with shallower indices (for e.g with $\epsilon=1$) are not consistent 
with the evolution described above.
\par

(ii) For our second example, we shall make an ansatz for the halo profile
and use it to determine the correlation function. 
We assume, based on small scale dynamics, that
the density profiles of individual halos 
should resemble that of isothermal spheres, with $\epsilon=2$, irrespective of 
initial conditions. Converting this halo profile to correlation function
in the {\it nonlinear} regime is straightforward and is based on equation
(\ref{gammep}):
If $\epsilon=2$, we must have $\gamma=2 \epsilon-3=1$ at 
small scales; that is $\bar\xi(a,x)\propto x^{-1}$ at the nonlinear regime.
Note that this corresponds to the critical index at the nonlinear
end, $n_{eff}=n_c=-2$ for which the growth rate is $a^2$ --- same as in linear
theory. 
[This is, however, possible for initial power law spectra, only if 
$\epsilon=1$, i.e $h(n+3)=1$ at very nonlinear scales.
Testing the conjecture that $h(n+3)$ is a constant is probably a little
easier than looking for invariant profiles in the simulations but the
results are still uncertain].

The corresponding analysis for the intermediate regime, with $1\gaprox\xb\gaprox 200$, is
more involved.
This is clearly
seen in equation (\ref{powcen1}) which shows that the power spectrum [and
hence the 
correlation
function] depends {\it both} on the fourier transform of the halo profiles as
well as the power spectrum of the distribution of halo centres. In general, 
both quantities will evolve with time and we cannot
ignore the effect of $P_{\rm cent}(k,a)$ and relate $P(k,a)$ to $f(k,a)$. 
The density profile around a {\it local maxima} will
scale approximately as $\rho\propto\xi$ while the density profile around
a {\it randomly} chosen point will scale as $\rho\propto\xi^{1/2}$. [The relation
$\gamma=2 \epsilon-3$   expresses the latter scaling of $\xi\propto\rho^2$]. 
There is, however,
reason to believe that the intermediate regime (with $1 \gaprox \bar\xi \gaprox 200$) is dominated by the
collapse of high peaks (\cite{TPMNRAS}) . In that case, we expect the
correlation function and the density profile to have the same slope
in the intermediate regime with $\xb\propto (1/x^2)$. Remarkably enough,
this corresponds to the critical index $n_{eff}=n_c=-1$ for the intermediate
regime for which the growth is proportional to $a^2$.

We thus see that if: (i) the individual halos are isothermal spheres
with $(1/x^2)$ profile and (ii) if $\xi\propto\rho$ in the intermediate regime
and $\xi\propto\rho^2$ in the nonlinear regime, we end up with a correlation
function {\it which grows as $a^2$ at all scales}. Such an evolution, of course,
preserves the shape and is a good candidate for the late stage evolution of
the clustering.

While the above arguments are suggestive, they are far from conclusive. It
is, however, clear from the above analysis and it is not easy to provide
{\it unique} theoretical reasoning regarding the shapes of the halos. 
The situation gets more complicated if we include the fact that all halos
will not all have the same mass, core radius etc and we have to modify our
equations by integrating over the abundance of halos with a given value of
mass, core radius etc. This brings in more ambiguities and depending on
the assumptions we make for each of these components [e.g, abundance for halos of a particular mass could be based on Press-Schecter or Peaks formalism],
and the final results have no real significance.
It is, therefore, better [and 
probably easier] to attack the question based on the evolution equation for
the correlation function rather than from ``physical'' arguments for density profiles. This is what we shall do next.

\section{ Self-similar evolution}

Since the above discussion motivates us to look for correlation functions
of the form $\xb=a^2L(x)$, we will start by asking a more general question:
Does equation (\ref{redefpair}) possess 
self-similar 
solutions of the form
\begin{equation}
\label{xiansatz}
\xb=a^{\beta}\:F(\frac{x}{a^{\alpha}})=a^{\beta} F(q)
\end{equation}
where $q\equiv x a^{-\alpha}$ ?. Defining $Q=\ln q=X -\alpha A$ and changing 
independent variables to from $(A,X)$ to $(A,Q)$  we can tranform our equation 
(\ref{redefpair}) to the form:
\begin{equation}
\left( \frac{\partial \bar \xi}{\partial A}\right)_{Q}-(h+\alpha) \left( 
\frac{\partial \bar \xi}{\partial Q}\right)_{A}=3 (1+\bar \xi)\:h(\bar \xi)
\end{equation}
Using the relations $({\partial \bar \xi}/{\partial A})_Q=\beta \bar \xi$,
 $({\partial \bar \xi}/{\partial Q})_{A}=(\bar \xi/F)(d F/d Q)$
we can rewrite this equation  as 
\begin{equation}
\label{KQeq}
\frac{\beta \bar \xi-3 (1+\bar \xi)h(\bar \xi)}{\left[\alpha+h(\bar 
\xi)\right]\bar \xi}=\frac{1}{F} \frac{d F}{d Q}\equiv K(Q)
\end{equation}
The right hand side of this equation depends only on $Q$ and hence will
vanish if differentiated with respect to $A$ at constant $Q$. Imposing this 
condition on
the left hand side and noticing that it is a function of $\xb$ we get
\begin{equation}
\left( \frac{\partial \bar \xi}{\partial A} \right)_Q \frac{d}{d \bar \xi}({\rm 
Left\:Hand\:Side})=0
\end{equation}
To satisfy this condition we either need (i) $(\partial \bar\xi/\partial 
A)_Q=\beta \bar\xi=0$ implying $ \beta=0 $ or (ii) the left hand side must be a 
constant.
Let us consider the two cases separately.

(i)
The simpler case corresponds to $ \beta=0 $ which implies that
$ \xb=F(Q) $. Setting $ \beta=0 $ in equation (\ref{KQeq}) we get

\begin{equation}
\left( \frac{d \bar \xi}{d Q} \right)=-\frac{3(1+\bar \xi)h(\bar 
\xi)}{\left[\alpha+h(\bar \xi)\right]}
\end{equation}
which can be integrated in a straightforward manner to give a relation
between $q=\exp Q$ and $\bar\xi$:
\begin{eqnarray*}
\label{qeq}
q&=&q_0 (1+\bar \xi)^{-1/3} \exp\left( -\frac{\alpha}{3} \int \frac{d \bar 
\xi}{(1+\bar \xi)h(\bar \xi)} \right)\\
&=&q_0 (1+\bar \xi)^{-1/3} {\cal V}(\bar\xi)^{-\alpha/2}
\end{eqnarray*}
Given the form of $h[\xb]$, this equation can be in principle
inverted to determine $\bar\xi$ as a function of $q=x a^{-\alpha}$.
\par
 To
understand when such a solution will exist, we should look at the limit
of  $\bar \xi\ll 1$. In this limit, when linear theory is valid,
we know that $h\approx (2/3) \bar \xi $ [see  \cite{Peeb80}]. Using this in 
equation
 (\ref{qeq})  we get the solution to be $\ln {\bar \xi}=-(2/\alpha) \ln q $ or
\begin{equation}
 {\bar \xi} \propto q^{-\frac{2}{\alpha}}\propto x^{-\frac{2}{\alpha}} a^2  
\propto a^2 x^{-(n+3)} 
\end{equation}
with the definition $\alpha\equiv 2/(n+3)$. This clearly shows that our solution 
is valid, {\it if and only if} the linear
correlation function  is a scale-free power law. In this case, of course, it is 
well known that solutions of the type $\bar\xi(a,x)=F(q)$ with $q=x 
a^{-\frac{2}{(n+3)}}$  exists. [Equation
 (\ref{qeq})  gives the explicit form of the function $F(q)$]. This result shows that this is the {\it only} possibility. It should be noted that, even though
we have no explicit length scale in the problem, the function $\bar\xi(q)$
--- determined by the above equation --- does exhibit different behaviour at
different scales of nonlinearity. Roughly speaking, the three regimes in
equation (\ref{Vapprox}) translates into nonlinear density contrasts in the ranges
$\delta<1,1<\delta<200 $ and $\delta >200$ and the function $\bar\xi(q)$
has different characteristics in these three regimes. This shows that
gravity can intrinsically select out a density contrast of $\delta\approx 200$
which, of course, is well-known from the study of spherical tophat collapse. 

(ii) Let us next consider the second possibility, {\it viz.} that the left hand
side of equation (\ref{KQeq}) is a constant. If the constant  is denoted by
$\mu$, then we get $F=F_0\:q^{\mu}$ and 
\begin{equation}
\beta\:\bar \xi-3\:(1+\bar \xi)\:h(\bar \xi)=\mu\:\alpha\:\bar \xi+\mu\:h\:\bar \xi
\end{equation}
which can be rearranged to give
\begin{equation}
\label{hform}
h=\frac{(\beta-\alpha \mu)\bar \xi}{3+(\mu+3)\bar \xi}
\end{equation}
This relation shows that solutions of the form 
$\bar\xi(a,x)=a^{\beta}\,F(x/a^{\alpha})$ with $\beta\neq 0$ is
possible only if $h[\xb]$ has a {\it very specific} form given by (\ref{hform}). 
In this form,
$h$ is a monotonically increasing function of $\xb$. There is, however,
firm theoretical and numerical evidence (\cite{Ham}; \cite{TPMNRAS})
to suggest that $h$ increases with
$\xb$ first, reaches a maximum and then decreases. In other words, the
$h$ for actual gravitational clustering is {\it not} in the form suggested by
equation (\ref{hform}). {\it We, therefore, conclude that solutions of the form 
in equation (\ref{xiansatz}) with $\beta\neq0$
cannot exist in gravitational clustering}.

By a similar analysis, we can prove a stronger result: There are no  solutions 
of the form $\bar \xi(a,x)=\bar \xi(x/F(a))$ except
when $F(a) \propto a^{\alpha}$. So self-similar
evolution in clustering is a very special situation.

This result, incidentally, has an important implication. It shows that power-law initial
conditions are very special in gravitational clustering and may not represent
generic behaviour. This is because, for power laws, we have a strong constraint
that the correlations etc can only depend on $q=x a^{-2/(n+3)}$. 
For more realistic --- non-power law --- initial conditions the shape can be
distorted in a generic way during evolution. 

All the discussion so far was related to finding {\it exact} scaling
solutions. It is however possible to find {\it approximate} scaling solutions
which are of practical interest. Note that we normally expect
constants like $\alpha,\, \beta,\, \mu$ etc  to be of order unity while $\xb$ can take arbitrarily
large values. If $\xb\gg 1$ then equation (\ref{hform}) shows that $h$ is 
approximately
a constant with $h=(\beta-\alpha\mu)/(\mu+3)$. In this case
\begin{equation}
\bar\xi(a,x)=a^{\beta} F(q) \propto a^{\beta} q^{\mu}\propto a^{(\beta-\alpha 
\mu)} x^{\mu} \propto a^{h(\mu+3)} x^{\mu}
\end{equation}
which has the form $\xb=a^{3h}F(a^hx)$ which was obtained earlier by
directly integrating equation (\ref{redefpair}) with constant $h$. We shall say 
more about such approximate solutions in the next section.

\section{Units of the nonlinear universe}

Having reached the  conclusion that {\it exact} solutions of
the form $\xb=a^2 G(x)$ are not possible, we will ask the question:
Are there such {\it approximate } solutions ? And if so, how do they
look like ? We will see  that such profiles --- which we shall call
``pseudo-linear profiles''--- that  evolve very close to the the above form
indeed exist.
In order to obtain such a solution and check its 
validity,
it is better to use the results of section 2.1  and proceed as
follows:

We are trying to find an approximate solution of the form 
$ \bar\xi(a,x)=a^2 G(x)$ 
to equation (\ref{redefpair}). Since the linear correlation function 
$\bar\xi_L(a,x)$ does grow as $a^2$ at fixed $x$, continuity demands that 
$\bar\xi(a,x)=\bar\xi_L(a,x)$ for all $a$ and $x$. [This can be proved more 
formally as follows: Let $\bar\xi=a^2 G(x)$ and $\bar\xi_L=a^2 G_1(x)$ for some 
range $x_1<x<x_2$. Consider a sufficiently early epoch $a=a_i$ at which all the 
scales in the range $(x_1,x_2)$ are described by linear theory so that 
$\bar\xi(a_i,x)=\bar\xi_L(a_i,x)$. It follows that $G_1(x)=G(x)$ for all 
$x_1<x<x_2$. Hence $\bar\xi(a,x)=\bar\xi_L(a,x)$ for all $a$ in $x_1<x<x_2$. By 
choosing $a_i$ sufficiently small, we can cover any range $(x_1,x_2)$. So 
$\bar\xi=\bar\xi_L$ for any arbitrary range. {\it QED}]. Since we have a formal 
relation (\ref{mapfun}) between nonlinear and linear correlation functions, we 
should be able to determine the form of $G(x)$. 

To do this we shall invert the form of the linear correlation function
 $\bar \xi_L(a,l)=a^2 G(l)$
and   write 
$l=G^{-1}(a^{-2} \bar \xi_L)\equiv F(a^{-2} \bar \xi_L)$
where $F$ is the inverse function of $G$. We also know that the linear correlation 
function $\bar\xi_L(a,l)$ at scale $l$ can be
expressed as ${\cal V}[\bar\xi(a,x)]$ in terms of the true correlation 
function $\bar\xi(a,x)$ at scale $x$
where 
\begin{equation}
\label{lx}
l=x (1+\bar \xi(a,x))^{1/3}
\end{equation}
So we can write
\begin{equation}
\label{lequ1}
l=F\left[\frac{\bar \xi_L(a,l)}{a^2}\right]=F\left[\frac{{\cal 
V}[\bar\xi(x,a)]}{a^2} \right]
\end{equation}
But $x$ can be expressed as $x=F[\bar\xi_L(a,x)/a^2]$;
Substituting this in (\ref{lx}) we have  
\begin{equation}
l=F\left[\frac{\bar \xi_L(a,x)}{a^2}\right]\left[
1+\bar \xi\right]^{1/3}
\end{equation}
From our  assumption $\bar\xi_L(a,x)=\bar\xi(a,x)$ ;
therefore this relation can also be written as
\begin{equation}
\label{lequ2}
l=F\left[\frac{\bar \xi(a,x)}{a^2}\right] \left( 1+\bar\xi \right)^{1/3}
\end{equation}
Equating the expressions for $l$ in (\ref{lequ1}) and (\ref{lequ2}) we get an 
implicit 
functional 
equation for $F$:
\begin{equation}
F\left[ \frac{{\cal V}[\bar \xi]}{a^2} \right]=F\left[ \frac{\bar \xi}{a^2} 
\right] \left(1+\bar\xi\right)^{1/3}
\end{equation}
which can be rewritten as
\begin{equation}
\label{feq}
\frac{F\left[{\cal V}(\bar \xi)/{a^2}\right]}{F\left[{\bar \xi}/{a^2}\right]}=
(1+\bar \xi)^{1/3}
\end{equation}
This equation should be satisfied by the function $F$ if we need to maintain the 
relation $\bar\xi(a,x)=\bar\xi_L(a,x)$. 
\par
To see what this implies, note that the left hand side 
 should not vary with $a$ at fixed $\bar \xi$. This is possible only if 
$F$ is a power law: 
\begin{equation}F(\bar \xi)=A \bar\xi^{m}\end{equation}
which in turn constrains the form of ${\cal V}(\bar \xi)$ to be 
\begin{eqnarray}
\label{Vform}
{\cal  V}(\bar \xi)=\bar \xi\:(1+\bar \xi)^{1/3m}
\end{eqnarray}
Knowing the particular form for ${\cal V}$ we can compute the corresponding 
$h(\bar\xi)$ from the relation
\begin{equation}
\frac{d \ln {\cal V}}{d \bar\xi}=\frac{2}{3} \;\frac{1}{(1+\bar\xi)\,h(\bar\xi)}
\end{equation}
For the ${\cal V}(\bar\xi)$ considered in equation (\ref{Vform}) we get
\begin{equation}
\label{hiform}
h=\frac{2 \bar\xi}{3+(3+1/m) \bar\xi}
\end{equation} 
which is the same result obtained by putting $\beta=2\;,\alpha=0$ in equation
(\ref{xiansatz}). We thus recover  our old result --- as we should --- that {\it exact\ } solutions 
of the form $\bar\xi(a,x)=\bar\xi_L(a,x)=a^2\;G(x)$ are {\it not} possible because the 
correct ${\cal V}(\bar\xi)$ and $h(\bar\xi)$ do not have the forms in 
equations (\ref{Vform}) and (\ref{hiform}) respectively. But, as in the
last section, we can look for approximate solutions.
\par
We note from equation (\ref{Vform}) that for $\bar\xi \gg 1$, we have
\begin{equation}
\label{anotherV}
{\cal V}(\bar\xi)={\bar\xi}^{(1+1/3m)};\qquad F(\bar\xi)\propto\bar\xi^m;\qquad 
G(\bar\xi) \propto \bar\xi^{1/m}
\end{equation}
This can be rewritten as 
\begin{equation}
\label{yetanotherv}
{\cal V}(\bar\xi)=\bar\xi^{\nu};\qquad 
F(\bar\xi)\propto\bar\xi^{1/3(\nu-1)};\qquad 
G(\bar\xi)\propto\bar\xi^{3/(\nu-1)}
\end{equation}
In other words if ${\cal V}(\bar\xi)$ can be approximated as $\bar\xi^{\nu}$, we 
have an approximate solution of the form
\begin{equation}
\label{approxsol}
\bar\xi(a,x)=a^2\;G(x)=a^2\;x^{3(\nu-1)}
\end{equation}
Since the ${\cal V}$ in equation (\ref{ham1}) is well  
approximated by the power laws in (\ref{Vapprox}) so that
\begin{eqnarray}
\label{approximateV}
{\cal V}(\bar\xi) &\propto & {\bar\xi}^{1/3} \;\;(1 \mbox{\gaprox} \bar\xi 
\mbox{\gaprox} 200)\\
&\propto&\bar\xi^{2/3} \;\;(200 \mbox{\gaprox} \bar\xi)
\end{eqnarray}
we can take $\nu=1/3$ in the intermediate regime and $\nu=2/3$ in the nonlinear regime. 
It follows from (\ref{yetanotherv}) that the approximate solution should have 
the form 
\begin{eqnarray}
\label{approxF}
F(\bar\xi) & \propto & {\frac{1}{\sqrt{\bar\xi}}} \;\;(1 {\mbox{\gaprox}} 
\bar\xi {\mbox{\gaprox}} 200)\\
& \propto & \frac{1}{\bar\xi}\;\;\;\;\;\;(200 \gaprox \bar\xi)
\end{eqnarray}
This gives the approximate form of a pseudo-linear profile which will grow
as $a^2$ at all scales.

There is another way of looking at this solution which is probably more physical and throws light on the scalings of pseudo-linear profiles. We recall that, in the study of finite gravitating systems made of point particles and
interacting via newtonian gravity, isothermal spheres play an important
role. They can be shown to be the local maxima of entropy [ see 
\cite{PadPhyRep}] and hence dynamical
evolution drives the system towards an $(1/x^2)$ profile. Since one expects
similar considerations to hold at small scales, during the late stages of evolution of the universe, we may hope that isothermal spheres with
$(1/x^2)$ profile may still play a role in the late stages of evolution of 
clustering in an expanding background. However, while converting the profile to correlation, we have to take note of the issues discussed in section 2.
In the intermediate regime, dominated by scale invariant radial collapse (\cite{TPMNRAS}), the density will scale as the correlation function and
we will have $\bar\xi\propto (1/x^2)$. On the other hand, in the nonlinear
end, we have the relation $\gamma=2\epsilon -3$ [see equation (\ref{gammep}) ] which
gives $\bar\xi\propto (1/x)$ for $\epsilon=2$. Thus, if isothermal spheres
are the generic contributors, then we expect the correlation function to
vary as $(1/x)$ and nonlinear scales, steepening to $(1/x^2)$ at intermediate
scales. Further, since isothermal spheres are local maxima of entropy, a configuration like this should remain undistorted for a long duration. This
argument suggests that a $\bar\xi$ which goes as $(1/x)$ at small scales
and $(1/x^2)$ at intermediate scales is likely to be a candidate for pseudo-linear profile. And we found that this is indeed the case.

To go from the scalings in two limits given by equation
(\ref{approxF}) to an actual profile, we can use
some fitting function. By making the fitting function sufficiently complicated,
we can make the pseudo-linear profile more exact. We shall, however, choose
the simplest interpolation between the two limits and try the ansatz:
\begin{equation}
\label{ansatzF}
F(z)=\frac{A}{\sqrt{z}\;(\sqrt{z}+B)}
\end{equation}
where $A$ and $B$ are constants. Using the original definition 
$l=F[\bar\xi_L/a^2]$ and the condition that $\bar\xi=\bar\xi_L$, we get
\begin{equation}
\frac{A}{\sqrt{\bar\xi/a^2}\;(\sqrt{\bar\xi/a^2}+B)}=l
\end{equation}
This relation implicitly fixes our pseudo-linear profile. Solving for $\bar\xi$, we get
\begin{equation}
\label{xisolution}
\bar\xi(a,x)=\left(\frac{Ba}{2}\;\left(\sqrt{1+\frac{L}{x}} -1\right)\right)^2
\end{equation}
with $L=4A/B^2$. Since this profile is not a pure power law, this will satisfy the equation (\ref{feq})
 only approximately. We choose $B$ such that the relation 
\begin{equation}
\label{justaneq}
F\left( \frac{{\cal V}(\bar\xi)}{a^2} \right) = F\left(\frac{\bar\xi}{a^2}
\right) 
\left(1+\bar\xi\right)^{1/3}
\end{equation}
is satisfied to greatest accuracy at $a=1$.

\begin{figure}[t]
\epsfxsize=300pt
\epsfysize=300pt
\epsfbox{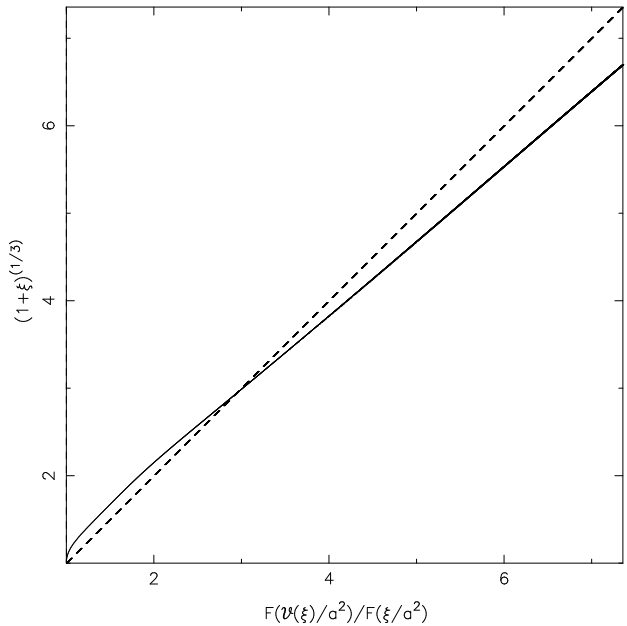}
\caption{The approximate solution to the functional equation
determining the pseudo-linear profile is plotted. See text for discussion.}
\end{figure}

This approximate profile works reasonably well. Figures 1 and 2 show this 
result. In figure 1 we have plotted the 
ratio 
$F( {\cal V}(\bar\xi)/{a^2})/F({\bar\xi}/{a^2})$ on the x-axis and
the function $(1+\bar\xi)^{1/3}$ on the y-axis. If the function in
(\ref{xisolution}) satifies equation (\ref{feq}) exactly, we should get a 45-degree line
in the figure which is shown by a dashed line. The fact that our curve is
pretty close to this line shows that the ansatz in (\ref{xisolution}) satisfies equation 
(\ref{feq}) fairly well. The optimum value of $B$ chosen for this figure is
$B=38.6$. When $a$ is varied from $1$ to $10^3$, the percentage of error
between the 45-degree line and our curve is less than about 20 percent in the
worst case.
 It is clear that our profile in (\ref{xisolution})  satisfies equation 
(\ref{justaneq}) quite well for a dynamic range of $10^6$ in $a^2$.

Figure 2 shows this result more directly. We evolve the pseudo linear profile 
form $a^2=1$ to $a^2\approx 1000$ using the NSR, and plot 
$[\bar\xi(a,x)/a^2]$ against $x$.
The dot-dashed, dashed and two solid curves (upper one for $a^2=100$ and lower one for $a^2=900$) are for $a^2=1,9,100$ and $900$
respectively. The overlap of the curves show that the profile does grow 
approximately as 
$a^2$. Also shown are lines of slope $-1$ (dotted)  and $-2$ (solid); clearly $\bar\xi\propto 
x^{-1}$ for small $x$ and $\bar\xi\propto x^{-2}$ in the intermediate regime.
\begin{figure}[p]
\epsfxsize=400pt
\epsfysize=400pt
\epsfbox{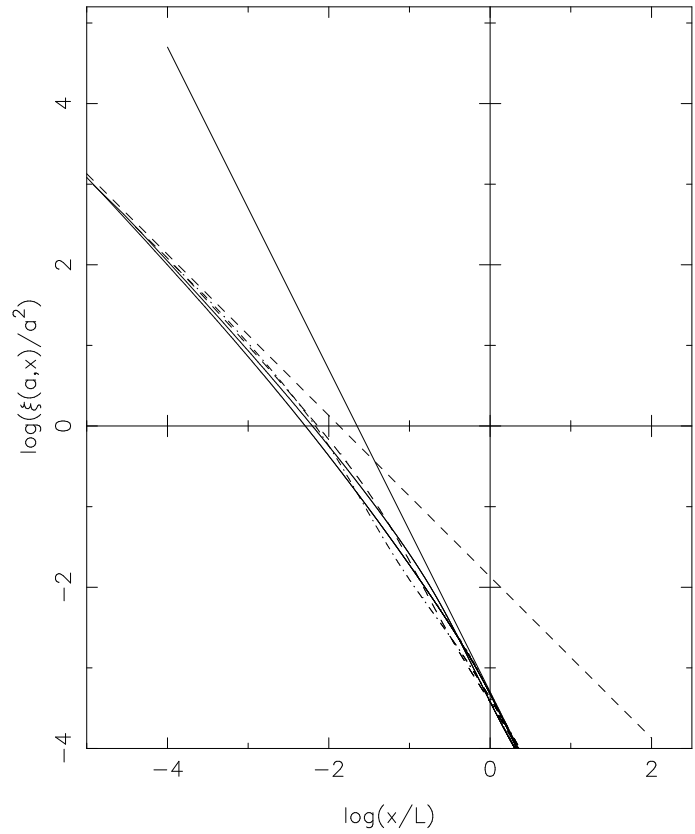}
\caption{The dot-dashed,dashed and two solid curves (upper one for $a^2=100$ and lower one for $a^2=900$) are for $a^2=1,9,100$ and $900$.
The dotted straight line is of slope -1 and the solid one is of slope-2 showing
both the $1/x$ and $1/x^2$ regions of the profile}
\end{figure}

We emphasis that we have chosen in equation (\ref{xisolution}) the simplest kind of ansatz combining the
two regimes and we have used only two parameters $A$ and $B$. It is quite
possible to come up with more elaborate fitting functions which will solve
our functional equation far more accurately but we have not bothered to do
so for two reasons: (i) Firstly, the fitting functions in equation (\ref{Vapprox}) 
for ${\cal V}(z)$ itself is approximate and is probably accurate only at
10-20 percent level. There has also been repeated claims in literature
that these functions have weaker dependence on $n$ which we have ignored for simplicity in this paper. (ii) Secondly, one must remember that only those
$\bar\xi$ which correspond to positive definite $P(k)$ are physically meaningful. This happens to be the case our choice [which can be verified by explicit numerical integration with a cutoff at large $x$] but this may not be true for arbitrarily complicated  fitting functions. Incidentally, another simple fitting function for the pseudo-linear profile is
\begin{equation}
\bar\xi(a,x)=a^2 \frac{A'}{(x/L')[(x/L')+ 1]}
\end{equation} 
with $A'=B^2$ and $L'=L/4$.

If a more accurate fitting is required, one can obtain it more directly from
equation (\ref{naeqn}). Setting $n_a=2$ in that equation predicts the instantaneous
spatial slope of $\xb$ to be
\begin{equation}
\part{\ln\xb}{\ln x}=\frac{2}{h[\xb]}-3(1+\frac{1}{\xb})
\end{equation}
which can be integrated to give
\begin{equation}
\ln\frac{x}{L}=\int_{\bar\xi[L]}^{\bar\xi[x]}\frac{h d\bar\xi}
{\bar\xi(2-3h)-3h}
\end{equation}
at $a=1$ with $L$ being an arbitratry integration constant. Numerical integration of this equation will give a profile which is
varies as $(1/x)$ at small scales and goes over to $(1/x^2)$ and then to
$(1/x^3),(1/x^4)....$ etc with an asymptotic logarithmic dependence. In the regime
$\xb>1$, this will give results reasonably close to our fitting function. 

It should be noted that equation (\ref{feq}) reduces to an identity
for any $F$, in the limit $\bar\xi\to 0$ since, in this limit $ {\cal V}(z)
\approx z$. This shows that we are free to modify our pseudo-linear profile
at large scales into any other form [essentially determined by the input 
linear power
spectrum] without affecting any of our conclusions.

Finally, we will discuss a different way of thinking about
pseudolinear profiles which may be useful.

In studying the evolution of the density contrast $\delta(a,{\bf x})$, it is 
conventional
to expand in in term of the plane wave modes as 
\begin{equation}
\label{name1}
\delta(a,{\bf x})=\sum_{\bf k} \delta(a,{\bf k}) \exp(i {\bf k}\cdot{\bf x})
\end{equation}
In that case,
the {\it exact} equation governing the evolution of $\delta(a,{\bf k})$  is 
given by (\cite{Peeb80})
\begin{equation}
\label{deltakeq}
\frac{d^2 \delta_{\bf k}}{d a^2}+\frac{3}{2 a} \frac{d \delta_{\bf k}}{d 
a}-\frac{3}{2 a^2}\delta_{\bf k}={\cal A}
\end{equation}
where ${\cal A}$ denotes the terms responsible for the 
nonlinear coupling between different
modes. The expansion in equation (\ref{name1}) is, of course, motivated by the 
fact that
in the linear regime we can ignore ${\cal A}$  and each of the modes evolve
independently. For the same reason, this expansion is not of much value
in the highly nonlinear regime.

This prompts one to ask the question: Is it possible to choose some other
set of basis functions $Q(\alpha,{\bf x})$, instead of $\exp\;i{\bf k}\cdot{\bf 
x}$, and expand $\delta(a,{\bf x})$ in the form 
\begin{equation}
\delta(a,{\bf x})=\sum_{\alpha} \delta_{\alpha}(a)\; Q(\alpha,{\bf x})
\end{equation}
so that the
nonlinear effects are minimised ? Here $\alpha$ stands for a set of parameters 
describing the basis functions. This question is extremely difficult to answer, 
partly because it is ill-posed. To make any progress, we have to first give 
meaning to the concept of ``minimising the effects of nonlinearity''. One 
possible approach we would like to suggest is the following: We know that when 
$\delta(a,{\bf x}) \ll 1 $,then $\delta(a,{\bf x})\propto a\:F({\bf x})$ for 
{\it any} arbitrary $F({\bf x})$; that is all power spectra grow as $a^2$ in the 
linear regime. In the intermediate and nonlinear regimes, no such general statement can 
be made. But it is conceivable that there exists certain {\it special} power 
spectra for which $P({\bf k},a)$ grows (at least approximately) as $a^2$ even in 
the nonlinear regime. For such a spectrum, the left hand side of 
(\ref{deltakeq}) vanishes (approximately); hence the right hand side should
also vanish. {\it Clearly, such power spectra are 
affected least by nonlinear effects.} Instead of looking for such a special 
$P(k,a)$ we can, equivalently look for a
particular form of $\xb$ which evolves as closely to the linear theory
as possible. Such correlation functions and corresponding power spectra [which 
are the pseudo-linear
profiles] must be capable of capturing most of the essence of nonlinear 
dynamics. In this sense, we can think of our pseudo-linear profiles as
the basic building blocks of the nonlinear universe. The fact that the
correlation function  is closely related to isothermal spheres, indicates
a connection between local gravitational dynamics and large scale gravitational
clustering.

\section{Conclusions}
It seems reasonable to hope that the late stage evolution of collisionless point particles, interacting via newtonian
gravity in an expanding background, should be understandable in terms of a simple paradigm. This paper [ as the title implies! ] tries to realise this dream
within some well defined framework. It should be viewed as a tentative
first step in a new direction which seems promising.

There are three key points which emerge from this analysis. The first is 
the fact that we have been able to find approximate correlation functions
which evolve preserving their shapes. We achieved this by looking at the
structure of an exact equation which obeys certain nonlinear scaling relations. As we emphasised before, the existence of such special class of solutions
to the equations of gravitational dynamics is an important feature.

Secondly, we should take note of the role
played by the ``isothermal'' profile $(1/x^2)$ in our solution. Such a profile can lead to
correlation functions which go as $(1/x)$ at small scales and $(1/x^2)$ in the
intermediate scales. If this profile is indeed ``special'' then one expects it
to lead to a  pseudo-linear profile for the correlation function. Our analysis shows that there
is indeed good evidence for this feature. If one accepts this evidence, then
the next level of enquiry would be to ask why $(1/x^2)$ profiles are ``special''.
In the statistical mechanics of gravitating systems, one can show that these
profiles arise as end stages of violent relaxation which operates at dynamical
time scales. Whether a similar reasoning holds in an expanding background,
independent of the index for power spectrum, is open to question. This is an
important issue and we hope to address it fully in a future work. We emphasise
that our equations, along with NSR, naturally lead to a pseudo-linear
profile, which can be interpreted and understood in terms of isothermal 
density profiles for halos; we did not have to assume anything a priori
regarding the halo profiles.

In a more pragmatic way, one can understand the pseudo-linear profile
from the dependence of the rate of growth of the correlation function
on the local slope. The NSR suggest that $\bar\xi$
grows (approximately) as $a^{6/(n_{eff}+4)}$ in the intermediate regime and as
$a^{6/(n_{eff}+5)}$ in the nonlinear regime. This scaling shows that $n_{eff}
=-1$ grows as $a^2$ in the intermediate regime and $n_{eff}=-2$ grows as $a^2$
in the nonlinear regime. This is precisely the form our pseudo-linear profile
has.  Also, in the intermediate regime, the correlation grows 
faster than $a^2$ 
if $n_{eff}<-1$ and slower than $a^2$ if $n_{eff}>-1$. The net effect is, of course, to straighten out a curved
correlation and drive it to $n=-1$. Similar effect drives the correlations
to $n=-2$ in the nonlinear regime.[see \cite{JsbTp} for a
more detailed discussion of this aspect in the intermediate regime]. Of course,
one still needs to understand the dependence of growth rate on the $n_{eff}$
from more physical considerations to get the complete picture. We have not addressed in this paper, what is the timescale over which clustering can lead to the psuedo-linear profile even granting that it does. This requires further study.

The last aspect has to do with what one can achieve using the pseudo-linear
profiles. In principle, one would like to build the nonlinear density field
through a superposition of pseudo-linear profiles but this is a mathematically
complex problem. As a first step one should understand why the nonlinear term
in equation (\ref{deltakeq}) is subdominant for such a profile. This itself is complicated
since we have only fixed the power spectrum --- but not the phases of the
density modes --- while the nonlinear terms do depend on the phase. Again,
we hope to investigate this issue further in a future work.

\acknowledgements
S.E thanks CSIR, India for financial support. Part of the work was done while
one of the authors (T.P) was visiting Astronomy Department, Caltech.
T.P thanks Caltech for hospitality.

\end{document}